\documentclass[11pt]{article}
\usepackage{hyperref}
\begin{document}
\title{Collision of a sphere onto a wall coated \\ with a liquid film}

\author{A.M. Ardekani and R.H. Rangel \\
\\\vspace{6pt} Department of Mechanical and Aerospace Engineering\\ University of
California, Irvine, CA 92697-3975, USA} \maketitle
%% The abstract (in this file, and that submitted as text to arXiv) should
%include the exact phrase
%% "fluid dynamics video" or "fluid dynamics videos"
\begin{abstract}
Particle-particle and particle-wall collisions occur in many natural
and industrial applications such as sedimentation, agglomeration,
and granular flows. To accurately predict the behavior of
particulate flows, fundamental knowledge of the mechanisms of a
single collision is required. In this fluid dynamics video,
particle-wall collisions onto a wall coated with 1.5\%
poly(ethylene-oxide) (PEO) (viscoelastic liquid) and 80\% Glycerol
and water (Newtonian liquid) are shown.
\end{abstract}
% main text
\section{Discussion}
In this \href{http://hdl.handle.net/1813/11394}{video}, the
collision of a sphere onto a wall coated with a 310 $\mu m$ film of
1.5\% PEO is shown. A string of liquid remains attached to the
particle even after a rebound height of 10$cm$. This illustrates the
extensional properties of the liquid. The wet coefficient of
restitution normalized with the dry one is 0.9$\pm$0.02 when the
Stokes number is 292 and the Weissenberg number is 0.21. The Stokes
number
is calculated using the film thickness as the length scale.\\

\noindent The normalized coefficient of restitution for a sphere
colliding onto a wall coated with a 538 $\mu m$ film of 80\%
Glycerol and water at Stokes number of 370 is 0.86$\pm$0.02. As
shown in the movie, a string of liquid remains attached to the
particle in the beginning, then it stretches 2.2$cm$ to later break
into several droplets. Bouncing motion of the droplets onto the
coated surface can be observed as well.

\end{document}